\documentclass[english,11pt]{article}
\usepackage[top=0.8in, bottom=0.78in, left=0.87in, right=0.87in]{geometry}
\usepackage{amssymb,amsmath,amsfonts,eurosym,geometry,ulem,color,setspace,sectsty,comment,footmisc,pdflscape,subfigure,array}
\usepackage{bm}
\usepackage[large, bf]{caption}

\usepackage{babel}
\usepackage{mathrsfs}
\usepackage{booktabs}
\usepackage{longtable}
\usepackage{nicefrac}
\usepackage{adjustbox}	% to adjust the size of objects to fit into a page
\usepackage[hyphens]{url}

\usepackage{breakurl}
\usepackage[pdftex,pagebackref=true,breaklinks]{hyperref}
\hypersetup{colorlinks, citecolor=blue, filecolor=blue, linkcolor=blue, urlcolor=blue}
\usepackage[authoryear,round,sort]{natbib}
\bibpunct{(}{)}{;}{a}{,}{,}

\usepackage{graphicx}
\usepackage{epstopdf}
\usepackage{titlesec}
\titlespacing*{\section}{0pt}{1.5ex plus 1ex minus .2ex}{0.8ex plus .2ex}
\titlespacing*{\subsection}{0pt}{1.2ex plus 1ex minus .2ex}{0.8ex plus .2ex}

\makeatletter
\g@addto@macro\@floatboxreset{\centering}
\makeatother

\usepackage{verbatim}

\graphicspath{{figures}}
\usepackage{soul}
\usepackage{epigraph}
\usepackage[]{epstopdf}
\makeatletter
\providecommand*{\input@path}{}
\g@addto@macro\input@path{{figures}}% append
\makeatother

\usepackage{dcolumn}
\newcolumntype{d}[0]{D{.}{.}{5}}

\title{\textbf{\LARGE{Cutting through Complexity: \\How Data Science Can Help Policymakers Understand the World}}\thanks{The views expressed in this chapter are those of the author, and not necessarily those of the Bank of England or its committees.}}

\author{
	Arthur Turrell\thanks{Bank of England. Email: \href{mailto:arthur.turrell@bankofengland.co.uk}{arthur.turrell@bankofengland.co.uk}\\
	This article will be a contributed chapter to the Santa Fe Institute edited volume: \textit{The Economy as a Complex Evolving System, Part IV}}
	}

\begin{document}

\maketitle
\thispagestyle{empty} 
\setcounter{page}{0}

\begin{abstract}\noindent
Economies are fundamentally complex and becoming more so, but the new discipline of data science---which combines programming, statistics, and domain knowledge---can help cut through that complexity, potentially with productivity benefits to boot. This chapter looks at examples of where innovations from data science are cutting through the complexities faced by policymakers in measurement, allocating resources, monitoring the natural world, making predictions, and more. These examples show the promise and potential of data science to aid policymakers, and point to where actions may be taken that would support further progress in this space.\\
\vspace{0in}
\vspace{0in}
\end{abstract}

%%%%%%%%%%%%%%%%%%%%%%%%%%%%%%%%%
% Text
%%%%%%%%%%%%%%%%%%%%%%%%%%%%%%%%%
\clearpage

\section{Introduction}

Economies are fundamentally complex and becoming more so, for a host of reasons. Declining response rates from surveys (once the bedrock of official statistics), the switch from tangible to intangible capital, mis-measurement of economic activity, the switch from production to services, the need to track different types of assets (for example, natural assets), increasingly specialised supply lines, and a demand for ever more granular solutions are just some of the headwinds facing policymakers as they try to navigate this complexity.

The challenge is made even greater as policymakers are now directly responsible for a greater share of gross domestic product (GDP) than ever before---for the thirteen countries in Europe for which the International Monetary Fund has long-run data, government expenditure as a percent of GDP grew by a mean of 26.6 percentage points between 1960 and 2022. And many of the noted challenges, especially with respect to measurement, are made harder in the public sector because of its diverse aims, and because the inputs, production, outputs, and outcomes are hard to ``count''.

Policymakers are particularly exposed to these complexities because they must try to look at the whole system, whether it be for statistical production, managing natural resources, determining who has what and why, or even deciding on who \textit{should} have what and why.

Data science provides powerful tools to chip away at this complexity. Some see this emerging field as a mishmash of other topics neglectfully crumpled together. The worry is that its adherents naively wade into other domains with impractical solutions. And that is a risk. But the examples in this chapter show that, at its best, data science can offer new ways to manage the explosion of information that policymakers are exposed to in the course of their decision making. And they show that, in addition to the existing precedents, numerous other high-impact applications lie just around the corner.

Others have called out data science's to aid policymaking. \citet{engin_algorithmic_2019} look at a number of case studies in what they call ``GovTech'', examining applications that could or have improved the workings of government. \citet{medeiros_data_2020} look more broadly, at businesses, and find that the increased agility with which insights may be obtained and the management of organisational performance are leading benefits of using data science. Studies have suggested that data science and the use of big data might improve productivity by as much as 7\% \citep{brynjolfsson_strength_2011,muller_effect_2018}. These productivity benefits are likely to come from automating processes, improving the quality of decision-making through better or more timely insights, and, via large language models, through providing aids to people---whether for writing, coding, or even tutoring for subjects like mathematics. \citet{bright_data_2019} examine the potential for UK local government to benefit from data science but find that severe budget constraints and a lack of appetite for innovative projects, especially those that could fail, have held back progress. A report \citep{europeancommission.jointresearchcentre._ai_2020} that brings together a number of examples of the use of artificial intelligence in public services in Europe is clear that there is limited evidence of large-scale benefits to date though it does highlight some useful early applications, including recognising whether agricultural land has been mowed in Estonia and predicting which child day care facilities it would be most efficient to inspect in Belgium.

\section{Data Science and What It Can Do}

Data science is a relatively new discipline that unites software engineering, analysis, statistics, domain knowledge, and the scientific method to leverage the maximum value from data of all types and sizes---from billions of rows of numerical data, to text, to images: any sort of recorded information that can be imagined. Its practitioners typically use code to do this and are proficient in one or more programming languages. Their combined knowledge of statistics, mathematics, algorithms, and coding allows them to extract value from data, for example in the form of insights, predictions, or even just smoother and more efficient operations. Predictions may be arrived at through the use of machine learning algorithms, and include the outputs from large language models and other types of generative artificial intelligence (AI). While data is the focus, the coding skills also mean that those trained in data science are able to automate digitised processes very effectively: for example, producing a report based on the latest data on a schedule without the need for human intervention. An additional skill of those trained in the discipline is communication, particularly of the insights from data and especially achieving this through visualisations---graphical representations of the data along with key messages derived from it. The best practitioners are comfortable communicating insights from data to technical and non-technical audiences alike. As an example, a firm might employ a data scientist to make sense of the success of marketing campaigns, to create dashboards of key performance indicators, or to run a machine learning algorithm that makes recommendations to its customers of what show to watch next. Public sector institutions might employ data scientists to provide rapid policy analysis, for example monitoring the positions and contents of ships during a supply chain crisis.

As a discipline, data science is often complementary to other topic areas. Not only does it draw on many other disciplines, it only reaches its full potential when integrated with a particular domain. As such, when it has been applied without considering the domain context, it has gone awry. There have been naive applications of machine learning to problems where the relationships between variables have shifted considerably---Google Flu Trends, which I will cover in more detail later in the chapter, is a classic example. Ignoring domain-specific context, not thinking about causality, and, relatedly, falling foul of the Lucas critique \citep{lucas_econometric_1976}, which says that we should not expect relationships in data to be invariant to policy or incentive changes, are all ways that data science applications can go wrong. In particular, data science is much more effective when its practitioners have a solid understanding of how data are being generated and the causal links behind that generating process. Bringing in deep domain expertise, either by collaborating with experts or by training data scientists in the domain, mitigates these risks and can unlock the full and enormous potential for problem solving that this new discipline offers.

As the field matures, data science is becoming smarter and data scientists are generating breakthroughs they can very much call their own. Causality plays an ever-larger role in the discipline, both in developing algorithms that combine causal inference with the power of data science algorithms (for example double machine learning as in \citet{chernozhukov_double_2018}), and in recognising that modelling the data-generating process is essential to reducing the amount of ``model shift'' \citep{kaur_modeling_2022}. A reduction in model shift means higher quality predictions for longer before a model has to be retrained on more recent data. As well as being smarter, some breakthroughs are recognisable as being primarily achievements in data science. Data visualisation is one such example: some of the authors of the widely used graphical plotting package \textbf{matplotlib} \citep{hunter_matplotlib_2007} developed entirely new colour schemes that are perceptually uniform to humans, and therefore more accurate in representing numerical information in charts \citep{nunez_optimizing_2018}. Generative artificial intelligence (AI) in general, and large language models in particular, are perhaps the most well-known major breakthroughs, and ones that have reached the general public with record speed. Protein folding predictions are another good example that, although based in a specific domain, could only have been possible using artificial intelligence. Proteins are molecules that are responsible for many of the processes fundamental to life at the cellular level, including converting food into energy and carrying oxygen around in blood. Predicting the three-dimensional structures of proteins based on the linear sequence of amino acids that they are made from is key to both understanding their role in life and in developing medical innovations, such as new drug targets, and had been a goal of biologists for decades. However, determining the three-dimensional structure of even a single protein previously took months to years of effort. A ground-breaking machine learning model called AlphaFold \citep{jumper_highly_2021} has a prediction accuracy competitive with experimentally determined protein structures, enormously cutting down the time and resources required to solve protein folding problems.

It's clear that data science can cut through complexity in multiple ways. For example, for understanding modern economies, the following attributes make data science useful:

\begin{itemize}
	\item It can turn unstructured information (text, images, audio) into tabular data, and those recorded data may ultimately be useful in measuring a phenomenon or aspect of activity. The YOLO (you only look once) machine learning algorithm for object detection is an extremely successful example of this; out of the box, it can detect and classify 80 objects from videos or still images \citep{wang_yolov9_2024}.
	\item It can navigate complexity between inputs and outputs with predictions. For example, to predict the weather over the next 10 days from current information historically required complex physics models, which were expensive because they needed to be run on supercomputers. Newer ``deep'' learning (a type of machine learning) models, however, are able to learn the mapping between conditions now and in 10 days' time, even though the weather is the canonical example of a chaotic system in which small changes in inputs can have vastly different outcomes \citep{lam_learning_2023}.
	\item It can programmatically deal with vast datasets, and complex networks. For example, in a single month, June 2023, UK citizens made 2.5 billion purchase transactions but modern data science tools can crunch through this and produce a range of analytical insights \citep{hoolohan_regional_23}. And, as another example, using network data science to capture the labour market as a series of related skills, rather than as mutually exclusive skills, can arguably give a better indication of which jobs and tasks are truly at risk of automation \citep{mealy_them_2022}.
	\item It can reduce high-dimensional information to a smaller amount of salient information. For example, in recommender systems---algorithms that might be used to suggest what media to consume next---the matrix of existing consumption of products by individual consumers might be very sparse. Dealing with large, sparse matrices is computationally costly and recommender systems that work directly with them perform poorly. To avoid the curse of dimensionality, reduction algorithms that capture a smaller number of latent factors can be used, and the recommendation problem solved on them. While singular value decomposition is often used in recommender systems, other algorithms that have found wide use for dimensional reduction include autoencoders \citep{tschannen_recent_2018} and uniform manifold approximation and projection \citep{mcinnes_umap_2020}.
\end{itemize}

\section{Cutting through Complexity}

\subsection{Cutting through Measurement Complexities}

What we measure tells the story, to some extent, of what we as a society care about. We have long measured economic activity because, imperfect as it is, it is consistently linked with better quality of life for most people. This interest hasn't changed, but the nature of economic activity has, and so the way we measure the economy must also change. But it's not just about what we produce or make available through services anymore, though that is a challenge. What we measure also reflects what we value---and, increasingly, we wish to make measurements of much that is not produced directly through economic activity: the natural world, happiness, even how we spend our time. And we care far more about making our measurements reflect distributions, as well as aggregates, now---albeit partly because modern computing makes this more possible than before. In all of these cases, our values are reflected and, if we are not careful, our biases too. For example, and as we shall see in more detail in this section, the inflation faced by those buying a budget basket of goods could be quite different from those buying an average basket of goods, and reflecting the experience of the former requires a different---and perhaps more difficult---measurement.

In this section, we shall see how data science can help with these measurement complexities. However, data science is no panacea---it can still be misled by the biases in, and values of, society. Biases can easily creep into machine learning models that are trained on the text that humanity has produced---if the original text tends to be misogynistic then unfortunately so will large language models trained on it, at least not without further intervention and correction. Another cautionary tale is that of the app through which residents can report finding a pothole in order to get their local government to fix it---which resulted in potholes mainly getting fixed in affluent areas where people could afford smartphones that would support the app \citep{crawford_hidden_2013}. Biases like these can, again, be corrected but they are a very real, if conquerable, risk for these applications of data science to measurement.

\subsubsection*{Measuring Complex Economies}

Large, advanced, service-based economies tend to be more complex to measure, understand, and improve than those based on agriculture and manufacturing because it is harder to count services, service improvements, and intangibles, than it is to count the rate of production of physical goods. Statistical agencies face numerous challenges \citep{luiten_survey_2020} in trying to understand what's happening with economic activity---especially given that response rates to surveys have declined by tens of percentage points since the 1970s \citep{stedman_end_2019}.

Data science enables new types of measurement, with greater granularity, to be made. As an example, at the UK's Office for National Statistics (ONS), prices have traditionally been collected by field agents visiting supermarkets with clipboards. This is an extremely reliable method of data collection, but the number of prices that can be obtained is necessarily limited because information is manually recorded by humans. In January 2022, the United Kingdom was experiencing rapid increases in the cost of living. Food campaigner Jack Monroe used social media to draw attention to the fact that price statistics as published (via the ONS's Consumer Price Index) did not describe the change in the \textit{lowest cost}, or budget, items that consumers might purchase in their grocery shop. This is true by design. This snippet from the ONS's website describes how to think about the aggregate price statistics: ``A convenient way to understand the nature of these statistics is to envisage a very large shopping basket comprising all the different kinds of goods and services bought by a typical household.'' But, Jack's point went, a barely-managing household is not a typical household.

Given the rapidly increasing public interest in the question of price increases for the poorest households, there was pressure for ONS to act. However, the traditional price data it collects was for the items purchased by the typical household only, and thus couldn't tell a comprehensive story about changes in the lowest prices.

Data scientists, led by myself and a colleague, were able to step in and web scrape price data from supermarkets' websites---thereby collecting information on the price of \textit{every} good available, including the lowest cost ones. Web scraping is rarely trivial, and was not in this case due to the need to extract product and price information from very differently structured websites. As well as legal agreements from the owners of the websites, the operation required advanced programming skills---and so could not have been done by statisticians alone. Furthermore, to construct a price index based on the lowest-cost items, the data scientists needed to be able to automatically create categories of products based on the names and descriptions of products, which can be achieved with data science techniques such as clustering, and track products in those categories over time. Additionally, these relatively large data needed to be organised and stored in a database. By bringing these skills together, data scientists at ONS were able to inform the public debate---and show that, indeed, the lowest available prices of some goods had increased by 20\% or 50\%, far higher than an equivalent ``typical'' grocery index of 6\% \citep{casey_tracking_2022}. This is an example of quality deepening---it provided a more granular economic measurement---but it's also an example of data science unlocking new insights. The data also showed that the \textit{next cheapest} product was often substantially more (20\%+) than the cheapest, making clear the importance of product line-up to those with constrained budgets.

Developments in data science can also deliver improvements in the accuracy and timeliness of statistics. The US Census Bureau's work on their construction indicator is a good example of this, as it's currently based on a survey that is expensive and takes time to compile. So they are also using machine learning on satellite images to classify whether construction on an approved site has started or not using known permit locations. To do this, they use a convolutional neural network trained to classify satellite images as having no construction yet, having started construction, or having completed construction. In the pilot, the accuracy of this approach was 92\% across all categories and there is a plan to extend it to other types of information that can be classified from satellite images. Given declining survey response rates, the other primary method of collecting these data, and the almost instantly available nature of satellite imagery, this approach delivers significant benefits. The Census Bureau worked with Statistics Canada on this project, and the latter has used the same approach to determine if construction is commercial or residential, and what crop types are under cultivation \citep{erman_use_2022}.

On numerous occasions, data science has helped to \textit{measure} the modern economy. In the United States, where driving to shops is common, satellite data has been used to estimate the extent to which individual retailers' stores have customers visiting them based on counts of parked cars near those stores \citep{bonelli_displaced_2023}. Data scientists have been quick to adopt so-called \textit{naturally occurring data} too, that is data that are generated naturally through the course of economic activity. These data are less directly suitable for plugging into statistical processes, like the compilation of national accounts, because they were never developed for that purpose. However, they cannot suffer from declining response rates as they reflect genuine economic activity. Vacancies as posted on online jobs boards are an example: the availability of these online job advertisements, and the fact that they constitute the vast majority of all job advertisements in the economy, means they can be transformed from a signal between a firm and prospective employees to a measure of labour demand \citep{turrell_transforming_2019}. They can potentially tell us about that demand in a much more granular way than surveys---for example, employers' attempts to attract women to jobs \citep{duchini_pay_2022} and the extent of flexible working arrangements \citep{draca_revolution_2022,adams-prassl_flexible_2020}. Many national statistical offices are considering how to incorporate other kinds of naturally occurring data into official statistics. Two examples that require data science skills are anonymised card payments, which constitute extremely big data (tens of millions of transactions a day in the United Kingdom) for national accounts and anonymised and partially aggregated mobile phone location data for internal and external migration.

\begin{center}
	\begin{figure}[ht]
		\includegraphics[width=\textwidth]{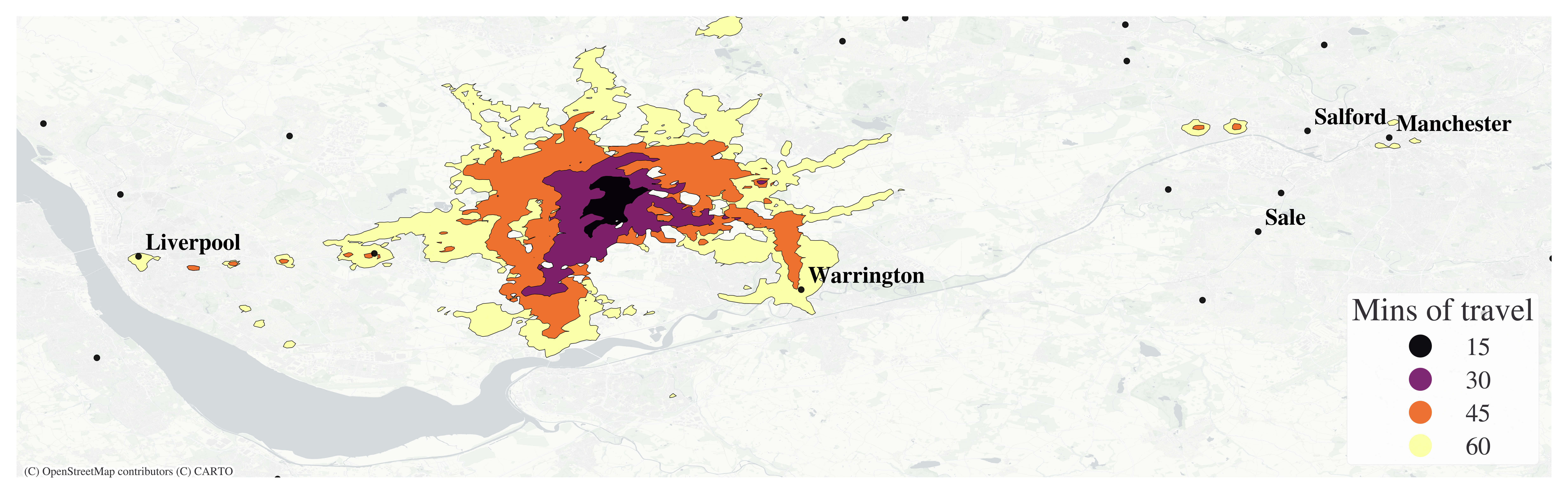}
		\caption{The feasible commuting area starting from St. Helens, UK, based on starting between 7:15 am and 9:15 am, travelling by a combination of foot, bus, and rail. \label{fig:travel}}
	\end{figure}
\end{center}

Services remain challenging for statistical agencies to measure. In the United Kingdom, the ability to access public and private services (for example hospitals, schools, good quality local shops, and good quality local restaurants) has been recognised by recent ``levelling up'' policies, and especially by a commitment to increase people's satisfaction with their local high street \citep{gove_levelling_22}. Similarly, good public transport opens up employment opportunities---a transport service that reliably delivers its passengers before the start of the working day can enable people to take jobs that an unreliable service cannot. Measuring the ability of the populace to access these kinds of in-person services is made difficult by the extremely heterogeneous means by which those living in different households could travel to them. However, modern data science techniques are able to handle both the complexity of the routing problem and the scale of the data required to analyse, hypothetically, how long it would take a person starting at arbitrary location A to get to an in-person service at location B. To help inform the debate about access to services, staff at the ONS were able to compute the area that people can feasibly travel to using public transport starting from hundreds of thousands of points across the United Kingdom---enabling anyone to identify groups of 40 to 150 households that might not be able to access services in nearby towns and cities \citep{banks_using_23}. An example of such a feasible commuting area is shown in Figure~\ref{fig:travel}. This work is a foundation for more granular access-to-services and access-to-jobs measures.

The complexity of the modern economy is present in goods too, as the vast range of products on offer today, and their subtle differences, can make it challenging to accurately record changes in prices once quality is taken into account. While hedonic quality adjustments have long existed \citep{goodman_andrew_1998}, the ability of data science and machine learning to capture unstructured data is enabling those subtle differences to be likewise captured, and also reduced from a vast amount of information into a single number (for example, a quality factor). This is a very direct example of both dimensional reduction and dealing with unstructured data. As an example, to help with Amazon's pricing strategies for clothing, \citet{bajari_hedonic_2023} developed hedonic models that can combine text, images, prices, and quantities, and output hedonic price estimates. To achieve this, they use machine learning models to generate arbitrary numerical factors that capture both text and images. The price predictions they are able to make with this model give a strong accuracy of over 80\%. The strength of this approach is that some features that distinguish the quality of goods may only be apparent to viewers of the item, and, previously, this has been entirely outside of the ability of economists or statisticians to capture or explain. Data science methods allow these features to be tracked and recorded. There is a lot of potential for this approach to be used to make quality adjustments in other contexts.

Complexity can arise in different ways. In some countries, it's less about a shift to services, and more about the difficulties of collecting data. Where countries have fewer resources to dedicate to the activities of their statistical offices, or conditions are dangerous because of a lack of infrastructure or the threat of conflict, data science can provide complementary statistics. As an example, since 2012, researchers have been using the night-time light emissions of developing countries as a proxy for GDP \citep{henderson_measuring_2012}, which is especially useful when official statistics are more difficult to collect or have a long lag. There is now a rich literature on using satellite data as a complement to traditional, survey-based statistics in developing countries where agriculture, which is discernible from space, tends to represent a larger fraction of economic activity.

\subsubsection*{Measuring the Natural World}

In an increasingly post-industrial world, what society cares about has come to encompass much more than providing the immediate needs for survival and, happily, has expanded to include preserving the natural environment as well as benefitting from it. Rapid environmental destruction and climate change have furthered the interest, and there is a strong need for statisticians and economists to turn their measuring tools to the natural world and its assets. This presents significant measurement challenges---it may be easy to count widgets coming out of a factory, but far harder to track the reduction in habitat of a moss. Data science provides peerless tools to measure, and even react to, the natural world.

There are a number of efforts that use remote sensors and computer vision---machine learning algorithms that can turn an image or video into counts of what is seen---to monitor the distribution of animals. One such example is \citet{teng_bird_2023}, who use citizen science-recorded bird location data to help train a model that uses satellite images to predict whether a range of bird species is present in an area. The model has some success at this prediction problem, with accuracies of more than 70\% for predicting which of the 684 species have been observed in a given region. Recent high-profile machine learning competitions have addressed the conservation of the Great Barrier Reef, of an endangered species of beluga whale, and of turtles by using facial recognition.

One complexity with Earth's changing climate is that the rapid change in the likelihood of dramatic and damaging events can disrupt well-functioning insurance markets because uncertainty makes pricing challenging. Wildfires, which are unfortunately becoming more frequent and more intense in US states like California, make this more pressing as they combine large population centres with conditions favourable to the ignition and rapid spread of fires. In 2018 alone, the damage from wildfires was equivalent to 1.5\% of the state's GDP. Efficient and timely prediction of where fires might break out is an important part of assessing risk. \citet{liu_simplified_2023} build a machine learning model that uses anthropogenic factors (such as power lines), topography, climate, and vegetation, and the US Forest Service's database of historical ignitions, to give annual estimates of the likelihood of ignition. Given the unfortunate fact of the increasing number of climate change-related events, models that are able to assess these kinds of risks will be increasingly useful.

Of course, climate change is also putting pressure on natural resources that humanity uses---water being a prime example. In 2022, the US Bureau of Reclamation sponsored a USD500,000 prize for estimating, in real time, the amount of water held in snow on mountains (which will eventually enter the water supply). The competition was particularly exciting because submissions were tested against data as it arrived in real time. The winning model explained 70\% of the variation in the water volume.

\subsection{Better Informed Decisions}

Although the examples of improved or new measurement are compelling in themselves, it's for no good if they do not actually improve the quality of decision-making. In this section, I present examples of where data science has cut through complexity to clearly and directly improve leaders' ability to make informed policy decisions.

\subsubsection*{Forecasting}

Forecasting is hardly new, but data science is bringing powerful new methods to bear on the problem, and these are, in relative terms, especially powerful in complex contexts. I already introduced one successful example of forecasting using data science, the weather; very much the canonical example \citep{lam_learning_2023}. But higher resolution, state-of-the-art, 2-hour ahead rain fore- and now-casts have also been achieved using machine learning \citep{ravuri_skilful_2021}.

There has been an explosion of new open source software tools for time series analysis and prediction coming out of data science in recent years, many of them developed by big technology firms, including Meta's Prophet \citep{taylor_forecasting_2017} and Kats, LinkedIn's Greykite \citep{hosseini_flexible_2021}, Uber's Orbit, Amazon's GluonTS, and Salesforce's Merlion.

The ability to incorporate non-numeric information and reduce the dimensionality of data to retain just the salient information has meant the development of new types of informative inputs for forecasts. For example, \citet{kalamara_making_2022} uses various methods to transform the extent to which the economy is being spoken about positively or negatively in newspapers into indicators. These signals, which they identify with animal spirits, are extremely real-time, and provide forecasting power for a range of macroeconomic variables above and beyond what can be gleaned from models that only incorporate traditional indicators. This kind of work provides a new and apparently informative measurement of economic sentiment. Another example is Google Trends. These data track the most popular searches of hundreds of millions of people and have been called a ``database of intentions''. There is little reason for an individual to perform a purposefully misleading search, so the combined searches of many individuals over time and in particular regions can be informative about macroeconomic trends \citep{choi_predicting_2012}. For example, one could use a search term related to claiming out of work benefits to anticipate rises in benefit claimants, and the Organisation for Economic Cooperation and Development (OECD) has incorporated the popularity of some search terms---such as ``luggage'' and ``bankruptcies''---into their real-time economic activity tracker \citep{woloszko_tracking_2020}.

It should also be noted that using these data for forecasting is not without pitfalls. Simply counting the occurrences of the words "financial crisis" from newspapers and putting that in a machine learning model would give a very misleading impression of the ability of text to forecast economic outcomes if applied today to the period 2007--2009. Back in 2007, you wouldn't have known to include that term in your analysis. Google Trends was famously used to track seasonal 'flu but, as a paper looking into why the model had predicted \textit{twice} the incidence of the disease put it, what was developed was part 'flu detector, part winter detector \citep{lazer_parable_2014}.

Regardless of the pitfalls, it seems clear that there are gains to be had from relatively easy-to-use and accurate forecasting tools. The potential in, say, local government, to predict seasonal demands on services based on a range of factors could play a big role in helping allocate resources more effectively.

\subsubsection*{Getting Information to Policymakers}

There are a range of topics where getting timely or detailed information to policymakers is likely to significantly improve the quality of decision making.

During the COVID-19 pandemic, there was a great need for high-frequency measures of human movement and mobility because mobility is a key determinant of how easily a virus spreads, and because existing official statistics of mobility were released with a substantial lag. ONS data scientists were challenged to develop faster indicators of within-country movement. In the United Kingdom, there are a large number of publicly owned and accessible closed circuit television (CCTV) feeds. So staff set up a cloud-based automatic data processing pipeline that woke up thousands of virtual computers every 10 minutes to draw down stills from CCTV cameras all over the country \citep{chen_estimating_2021}. Without direct oversight from any human, these images were then run through an anonymisation algorithm that blurred any faces or vehicle number plates that could be used to identify individuals. Next, the images were sent through a region-based recurrent neural network \citep{ren_faster_2015}, a type of machine learning algorithm. This was trained to recognise pedestrians, cars, vans, and cyclists and create counts of each of them. With so many different cameras in play, issues such as broken feeds, artefacts on screenshots, and dropped connections were common, so there was some post-count imputation of missing values. Finally, the mobility counts were aggregated into type and region.

\begin{center}
	\begin{figure}[h]
		\includegraphics[width=0.5\textwidth]{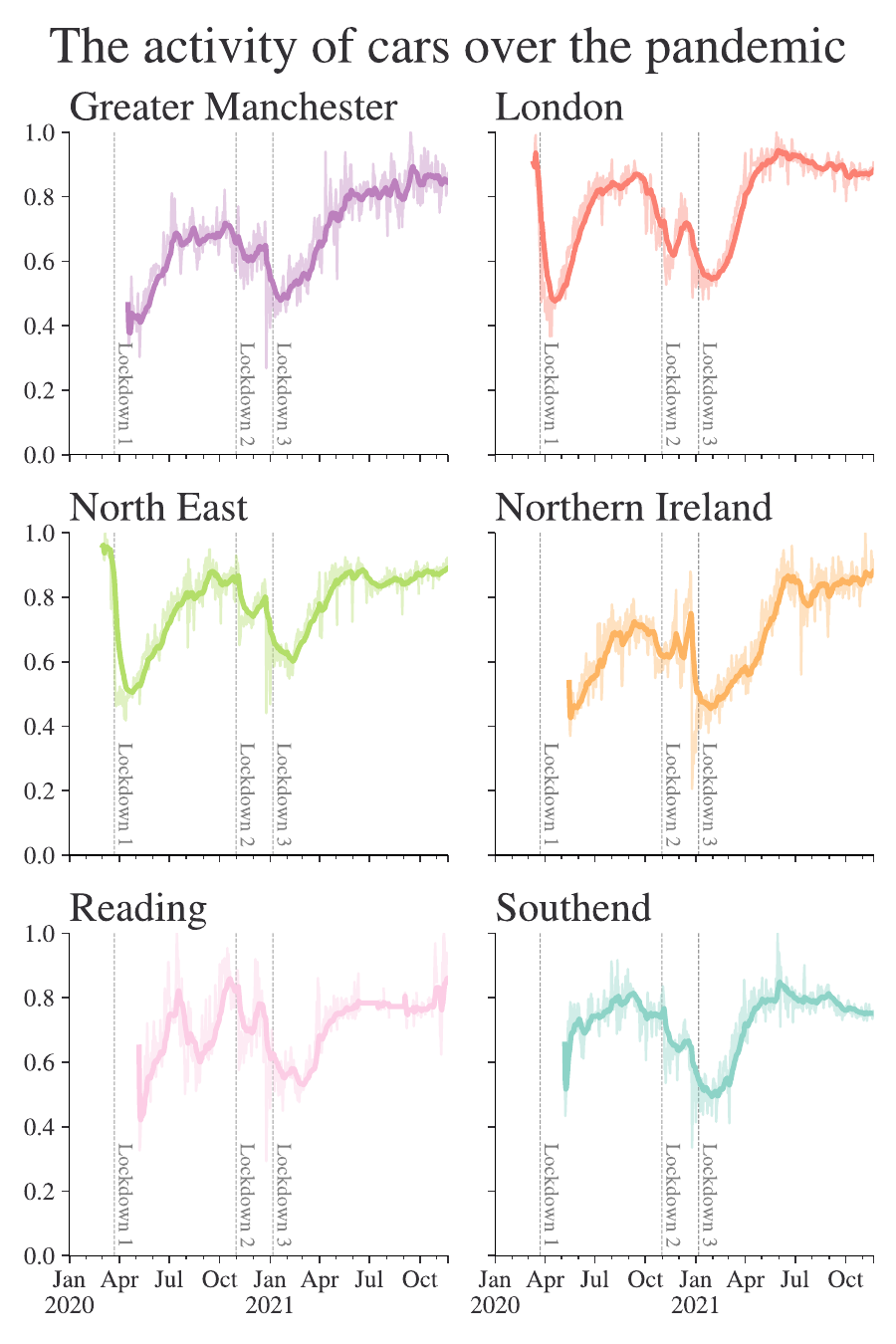}
		\caption{Activity of cars based on piping still frames from publicly accessible CCTV cameras to a series of machine learning algorithms; see \citet{chen_estimating_2021} for more information. The anticipatory effects of lockdowns are apparent. \label{fig:cars}}
	\end{figure}
\end{center}

This new monitoring system, fusing unstructured data and existing public assets, was able to deliver real-time estimates of the extent to which people were moving around the country---and inform policymakers whether or not additional mobility-related measures were necessary. Figure~\ref{fig:cars} shows some results from this algorithm, and in particular that not only did lockdowns severely reduce mobility by car, but also that there were anticipatory effects that saw the reduction begin before policy interventions were fully in place. This pipeline is estimated to have cost ONS around GBP20 a day to run.

Supply chain issues were also a prominent feature of the COVID-19 pandemic, and it became clear that monitoring systems that could let policymakers know what goods might be lacking in the near future would be extremely useful for informing actions. Few goods are as critical as food, and many countries rely on imports to meet a substantial portion of their needs. In emergencies, having accurate and timely information on food supply and distribution is critical to reducing human suffering. \citet{balashankar_predicting_2023} took a new approach to providing information that was more timely in predicting when food crises would occur in countries with a history of food insecurity. They developed a food crisis early warning system that uses newspaper articles published in at-risk countries to give as much as 12 months' notice of food insecurity events. The model is trained on 11 million news articles and outperforms expert predictions alone (and, as is typical, combining expert predictions with the machine learning predictions yields \textit{significantly} improved performance). Although there will likely be some feedback effects to contend with, the approach demonstrates how data science can lend decision-makers extra foresight and time to act.

Perhaps the most urgent occasions when policymakers must make decisions quickly are when natural disasters hit. This can be especially challenging in countries where there are fewer resources available to go around when a disaster occurs. Earthquakes are a good example of where data science, rapidly deployed, can help assess where, and how badly, building damage has occurred in order to distribute rescue parties and other first-responders effectively. Approaches that have been introduced include using geographically precise data on earthquake intensity alongside detailed building characteristics (such as number of stories and age) to predict building damage \citep{mangalathu_classifying_2020} in Nepal, and using pre- and post-earthquake satellite images, along with vision (in this case YOLO) and segmentation models, to quickly label damaged buildings on maps that were sent to first responders in Turkey.

\section{Summary and Conclusion}

The examples presented in this chapter show data science's great potential to help policymakers. But it's likely these are only the start of the story---a 2019 report found that few UK local government organisations were using any machine learning in production \citep{bright_data_2019}. It seems clear that this relatively new discipline could provide a whole host of additional improvements, many yet to be imagined. Because of its ability to make processes more efficient and deliver new insights, data science may provide a path to improving public-sector productivity, and this is significant given the size of the public sector in many economies.

For data science to continue to improve policymakers' understanding of the world, the conditions need to be right. Much of data science is built on free and open source software, which, in aggregate, is estimated to be worth USD8.8 million million, which is what it could cost to replace it \citep{hoffmann_value_2024}. But for free and open-source software to thrive, workers at firms, in the public sector, and in academia, need to be given the right incentives to produce it.

Data are also important: as a former colleague put it, ``You can't data science without data,'' and there are challenges with regards to this input to the data science production function. While, historically, statistical offices had access to the data (typically surveys) that went into creating national accounts, the measurement challenges of today, and the need for greater granularity, mean that some key data sources are in private hands. For example, for a real-time view of consumption in nations with large economies, there are few data sources better than credit and debit card data---but the payment systems that track these data are usually privately held, and require national statistical offices to either buy the data or enter into complicated legal agreements to access them. Such data need careful treatment to protect privacy too, though there's nothing new in that. There are great success stories where private data has been acquired, but it involves much more co-ordination, planning, and goodwill than running a survey. The difficulty in setting up these agreements affects the rate of progress---while national statistical offices may have the purchasing power to buy these datasets, or can rely on legislation to acquire them, academics who would ideally be working on improving how these data, say, map into official statistics typically have neither, and no access. We know that when datasets have been made public (in the case that they do not contain sensitive data), they deliver huge value, \citet{loomis_valuing_2015} show that the US government releasing Landsat images without cost resulted in a net efficiency gain worth tens of millions of USD per year. To keep providing improvements through data science, it's likely that the public sector will need to find ways to allow the wider community to connect to, and develop on, its sensitive datasets securely, either through cloud platforms or the provision of synthetic data.

And acquiring \textit{some} data isn't always enough. Machine learning models scale aggressively with compute and data ingested---but most institutions do not have access to the resources required to feed in enough data to create the most powerful models. It's a problem that is likely to get worse as the early providers of, say, large language models, have a first-mover advantage: they can train on their customers' queries and so bootstrap the performance of their models in a way that is hard for others to replicate. Potential solutions include nationally supported foundation models and government supported co-ordination on improving a small number of the most promising open-source models.

Data is not the only input. Computing capital matters enormously too, and institutions need to provide the right tooling for data scientists to do their work. Today, many firms and public sector organisations have legacy IT that greatly limits what is possible, or do not sufficiently empower workers to use computing services effectively. It is typical for a person with a laptop and a cloud computing account acting privately to have more flexibility, power, choice, and, ultimately, productivity in their data science tooling than an individual using a large institution's locked-down and centralised IT platform. But, of course, the most useful data for public policy analysis tend to only be available on the latter, so to get the best outputs we must bring together the best inputs (data) and the best capital (tooling). Fortunately, due to Amazon Web Services, Google Cloud, Azure, and other providers, there is a clear path for organisations to get the best of their institutional data and provide best-in-class tooling, but it does require them to empower workers with comprehensive access to these cloud services.

Despite these challenges, the examples in this chapter show the potential of data science to help policymakers. The discipline offers a compelling set of tools for cutting through the complexities of measurement and decision-making, and holds out the prospect of productivity improvements to boot.

\newpage
\singlespacing
\bibliographystyle{aeamod}
\bibliography{bibliography_clean}

\end{document}